# Multiobjective decomposition of integer matrices: application to radiotherapy

T. Lust *, J. Teghem

*University of Mons*
*UMONS - Laboratory of Mathematics and Operational Research*
*20 Place du Parc, 7000 Mons (Belgium)*

**Abstract**

We consider the following problem: to decompose a nonnegative integer matrix into a linear combination of binary matrices that respect the consecutive ones property. This problem occurs in the radiotherapy treatment of cancer. The nonnegative integer matrix corresponds to fields giving the different radiation beams that a linear accelerator has to send throughout the body of a patient. Due to the inhomogeneous dose levels, leaves from a multi-leaf collimator are used between the accelerator and the body of the patient to block the radiations. The leaves positions can be represented by segments, that are binary matrices with the consecutive ones property. The aim is to find efficient decompositions that simultaneously minimize the irradiation time, the cardinality of the decomposition and the setup-time to configure the multi-leaf collimator at each step of the decomposition. We propose for this $\mathcal{NP}$-hard multiobjective combinatorial problem a heuristic, based on the adaptation of the two-phase Pareto local search. Experiments are carried out on different size instances and the results are reported.

## 1 Introduction

In 1896, Wilhelm Conrad Roentgen (who received the first Nobel prize in physics in 1901) reported his discovery of the x-ray (called "x-ray" by Roentgen because it was an unknown type of radiation at this time). It is a form of electromagnetic radiation than can cause cell damage by interfering with the ability of the cells to grow and reproduce [19]. Few months later, along with the parallel discovery of radium by Marie Curie (Nobel prize in physics in 1903), the first cancer patient was irradiated [21]. The use of radiotherapy is based on the fact that cells especially sensitive to radiation are those that reproduce more frequently, such as cancer cells. On the contrary, even if they are also affected by radiation, normal cells possess the ability to repair and recover from radiation damage. Following this understanding, the medical community adapted radiotherapy to destroy cancer cells. More than 100 years later, radiotherapy is used by more of the half of people with cancer [24]. Radiotherapy can also be combined with surgery or chemotherapy. Its use depends on the type of the tumor, on the localization of the tumor and on the health of the patient.

* Corresponding author. Address: thibaut.lust@umons.ac.be.

The radiotherapy treatment works as follows. A patient lies down on a couch and a linear accelerator, mounted on a gantry, sends radiations in the direction of the tumor in order to kill it (see for example the figure 1 in [10]). The radiations are composed of beams of high energy photons or electrons. The treatment should be such that enough dose is delivered to the targeted region, to kill the cancerous cells, while preserving the surrounding anatomical structures. It is also possible to move the gantry or the couch to treat the tumor from various angles, in order to prescribe a dose across all three dimensions so that the amount of radiations sent to the healthy organs surrounded the tumor is balanced.

Nowadays, intensity-modulated radiation therapy (IMRT) is mainly used (the first clinical IMRT with modern technology was in March 1994 [26]). The IMRT is an advanced type of high-precision radiation, due to the advancements during the 80's and 90's in computer technology, electronic miniaturization, three-dimensional digital imaging and specialized software tools [21]. IMRT allows to optimally conforms radiation beams to tumor shapes thanks to three-dimensional imaging and a device called multileaf collimator (MLC).

The MLC is composed of a set of leaves, made of usually tungsten, that can move independently in and out of the path of a beam in order to block it [1]. The role of the MLC is to modulate the radiations. Indeed, by combining different configurations of the MLC, it is possible to obtain different intensities to reduce radiation doses delivered to the surrounding healthy organs and tissues.

Some advantages of IMRT are the following [24]:

- Fewer side effects and higher and more effective radiation doses.
- Possibility to treat concave organs.
- Better dose repartition.

---

[1] See for example the Varian medical system at `http://www.varian.com`.

- Cure rates in cancer patients is improved by 15-40 percent in comparison with conventional radiotherapy.

On the other hand, IMRT makes the treatment more complex and demands new skills for the physician, dosimetrist and physicist. It can take as long as 30 minutes to give an IMRT treatment, versus 5-10 minutes for conventional radiotherapy. Also, a radiation oncology department may need to invest millions (about 7 millions EUR for a new operational unit [24]) to get a new IMRT machine, planning software, and increased personnel. Furthermore, it takes more manpower to design a radiation plan and to deliver the treatment.

Treatment planning is achieved in most systems using inverse planning software algorithms. First, imaging techniques (computed tomography, magnetic resonance imaging, positron emission tomography,...) are used to diagnose and localize tumors. Images of both target tissues and tissues at risk are required to define upper bounds doses to healthy organs and lower bounds doses to the target. Then the IMRT optimization process determines the beam parameters that will lead to the desired solution.

The IMRT optimization process is usually composed of three phases [7]:

- The selection of the number of beams and the beam angles through which radiation is delivered (geometry problem);
- The computation of an optimal intensity map for each selected beam angle (intensity problem);
- The determination of a sequence of configurations of the MLC (realization problem).

Even if these problems are connected, they are often solved independently (that can however only gives suboptimal overall results).

The geometry and intensity problems are very closed. In these problems, the aim is to find locations of beams and beam intensities in order to achieve a high uniform dose in the target volume to increase tumor control, while limiting the dose to healthy organs and tissues as much as possible to reduce the probability of complications (see for example the figure 1 in [2]).

There exist multiobjective formulations of these problems, since both criteria, that is high radiation in target volume and low radiation in organs at risk are conflicting. For a review of the literature of these two problems, we refer the reader to the recent survey of Ehrgott *et al.* [7].

We only consider in this work the realization problem. We thus make the assumption that the beam angles are fixed and that for each beam angle, the intensity matrix is given (which is not constant since inhomogeneous dose levels are administrated: certain cancer targets receive a required amount of dose while functional organs are spared).

We present in the next section the mathematical formulation of the realization problem, that is how to realize the intensity matrix with the MLC.

## 2 The realization problem

Throughout we use the notation $[n] := \{1, 2, \cdots, n\}$ for positive integers $n$.

We consider a nonnegative integer matrix $A$ of size $m \times n$: $A = (a_{i,j})$ with $i \in [m]$ and $j \in [n]$. The matrix corresponds to the intensity matrix giving the different values of radiation that the linear accelerator sends throughout the body of a patient. The value $a_{i,j}$ of $A$ gives the desired intensity that should be delivered to the coordinate $(i, j)$.

We have to decompose the matrix $A$ into a set of segments. The segments correspond to the shape of the MLC. In the MLC, we make the distinction between the leaves located on the left (left leaves) and those located on the right (right leaves).

A segment can be represented by a special binary matrix of size $m \times n$ that describes the left and right leaves positions. These matrices have to respect the consecutive ones property (C1), which means, in short, that the ones occur consecutively in a single block in each row (since we can only block the radiations with a left or a right leaf).

A segment is noted $S = (s_{i,j})$ with $i \in [m]$ and $j \in [n]$. An example of a decomposition of a matrix $A$ into segments is shown below.

$$
A = \begin{pmatrix} 4 & 8 & 3 \\ 5 & 2 & 1 \\ 5 & 7 & 2 \end{pmatrix}
$$

$$
= 4 \begin{pmatrix} 1 & 1 & 0 \\ 1 & 0 & 0 \\ 1 & 1 & 0 \end{pmatrix} + 2 \begin{pmatrix} 0 & 1 & 1 \\ 0 & 1 & 0 \\ 0 & 1 & 1 \end{pmatrix} + 1 \begin{pmatrix} 0 & 1 & 1 \\ 1 & 0 & 0 \\ 1 & 1 & 0 \end{pmatrix} + 1 \begin{pmatrix} 0 & 1 & 0 \\ 0 & 0 & 1 \\ 0 & 0 & 0 \end{pmatrix}
$$

The positions of the left and right leaves corresponding to a segment $S$ are given by the $l_i$ and $r_i$ integers defined as follows:

$$0 \leq l_i < r_i \leq n+1 \quad (i \in [m])$$

$$s_{i,j} = \begin{cases} 1 \text{ if } l_i < j < r_i & (i \in [m], j \in [n]) \\ 0 \text{ otherwise.} \end{cases}$$

We denote the set of feasible segments by $\mathcal{S}$.

For example, for the following segment:

$$\begin{pmatrix} 1 & 1 & 0 \\ 1 & 0 & 0 \\ 0 & 0 & 1 \end{pmatrix}$$

we have: $l_1 = 0, r_1 = 3; l_2 = 0, r_2 = 2$ and $l_3 = 2, r_3 = 4$.

It should be noted that we consider that a line full of zero is arbitrarily defined by $l = 0$ and $r = 1$.

A feasible decomposition of $A$ is a linear sum of segments and has the following form:

$$A = \sum_{k=1}^{K} u_k S^k \text{ with } u_k \in \mathbb{N}_0, S^k \in \mathcal{S}, \ \forall k \in [K].$$

Two criteria are generally considered to evaluate the quality of a decomposition: the total irradiation time and the setup-time.

The total irradiation time, crucial from a medical point of view, is the time during which a patient is irradiated. This criterion is proportional to the sum of the coefficients (**decomposition time**).

The setup-time is the time needed to configure the MLC. This criterion is generally considered as proportional to the number of segments (**decomposition cardinality**). It is important to minimize this criterion in order to reduce the time of the session and so the comfort of the patient. Moreover, an optimal scheduling for radiotherapy is more and more required since the number of patients treated by radiotherapy increases. It is only recently that this scheduling problem has caught the attention of the operations research community [6, 15]. Obviously, minimizing the time of the session will improve the quality of the scheduling and the number of patients treated.

We can formulate both objectives that is the decomposition time (DT) and the decomposition cardinality (DC) as follows:

$$(\textbf{DT}) \quad \min \left\{ \sum_{k=1}^{K} u_k \ \Big| \ A = \sum_{k=1}^{K} u_k S^k, u_k \in \mathbb{N}_0, S^k \in \mathcal{S}, \ \forall\, k \in [K] \right\}$$

$$(\textbf{DC}) \quad \min \left\{ K \ \Big| \ A = \sum_{k=1}^{K} u_k S^k, u_k \in \mathbb{N}_0, S^k \in \mathcal{S}, \ \forall\, k \in [K] \right\}$$

Polynomial algorithms are known for the DT minimization [4, 22]. Many optimal solutions can be found for this single-objective problem.

On the other hand, the DC minimization has been proved to be $\mathcal{NP}$-Hard [3, 5].

Some authors [3, 9, 14] optimize the objectives lexicographically: by first minimizing DT and then trying to reduce DC while keeping the minimal value for DT. Taşkin *et al.* [23] and Wake *et al.* [25] recently considered both objectives at the same time, but by simply doing a linear sum of the objectives.

To be more realistic, we do not only consider constant times to move from one segment to the next but variable times depending on the time to move from one configuration of the MLC to another. The **variable setup-time** is defined as follows:

$$(\textbf{SU}_{var}) \quad \min \left\{ \sum_{k=1}^{K-1} \mu(S^k, S^{k+1}) \ \Big| \ A = \sum_{k=1}^{K} u_k S^k, u_k \in \mathbb{N}_0, S^k \in \mathcal{S}, \ \forall\, k \in [K] \right\}$$

where $\mu$ is proportional to the time necessary to change the setup of the MLC from the configuration corresponding to $S^k$ to the configuration corresponding to $S^{k+1}$. This objective is also known under the name overall leaf travel time [14]. The overall leaf travel time objective does not have to be neglected since the velocity of the leaves is limited due to precision issues. Moreover, the more the leaves have to move, higher the risk of imprecision is [27].

The value $\mu$ between two segments $S^k$ and $S^{k+1}$ is computed as follows [8]:

$$\mu\left(S^k, S^{k+1}\right) = \max_{1 \leq i \leq m} \max \left\{ \left| l_i^{k+1} - l_i^k \right|, \left| r_i^{k+1} - r_i^k \right| \right\}$$

It thus corresponds to the maximal move that a leaf has to make to attain its next position. Between two different segments, the minimal value that can take $\mu$ is equal to one.

For the example below, where we have to go from $S^1$ to $S^2$, we have: $\mu(S^1, S^2) = \max\Big( \max(|2-0|, |4-4|), \max(|0-1|, |3-3|) \Big) = 2$.

$$S^1 = \begin{pmatrix} 1\ 1\ 1 \\ 0\ 1\ 0 \end{pmatrix} \begin{bmatrix} l_1 = 0,\ r_1 = 4 \\ l_2 = 1,\ r_2 = 3 \end{bmatrix} S^2 = \begin{pmatrix} 0\ 0\ 1 \\ 1\ 1\ 0 \end{pmatrix} \begin{bmatrix} l_1 = 2,\ r_1 = 4 \\ l_2 = 0,\ r_2 = 3 \end{bmatrix}$$

Once the segments are fixed, the minimization of $\text{SU}_{var}$ is equivalent to a search for a Hamiltonian *path* of minimal cost on the complete graph which has the segments as vertices and the cost function $\mu$ on the edges (see Fig. 1). The cost function $\mu$ has the property to be a metric [14].

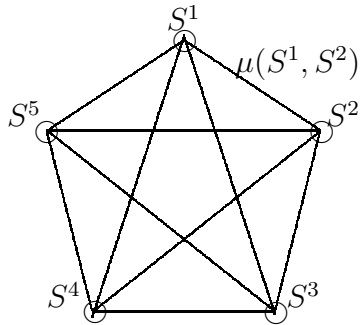

Fig. 1. Minimization of the overall leaf travel time by searching a Hamiltonian path of minimal cost on the complete graph which has the segments as vertices and the cost function $\mu$ on the edges.

This problem can be easily transformed to a TSP problem (Hamiltonian *cycle* of minimal cost) by adding a dummy vertex which has a distance of zero to all other vertices.

The key of the minimization of $\text{SU}_{var}$ is in the generation of segments that give a solution with a Hamiltonian path of minimal value, since once the segments are fixed, the minimization of $\text{SU}_{var}$ is relatively easy (if the number of segments remains low).

By considering the $\text{SU}_{var}$ objective, we also remark that modifying the sequence of segments has an impact on the value of $\text{SU}_{var}$, while this modification has no impact on DT and DC.

As there is a positive correlation between DC and $\text{SU}_{var}$, the few authors that have considered $\text{SU}_{var}$ tried to first minimize DC and then $\text{SU}_{var}$ by generating the best sequence of segments. Kalinowski [14] used a minimum spanning tree approximation to find the best sequence of segments, but it also possible to use an exact TSP algorithm, as done by Ehrgott *et al.* [8]. Siochi [22] considered both objectives at the same time, but through a linear sum.

In this work, in order to get a full overview of the relations between the objec-tives, we consider the three objectives (DT, DC, SU$_{var}$) simultaneously. The multiobjective formulation of the multiobjective combinatorial optimization (MOCO) problem (P) considered is thus the following:

$$(P) \begin{cases} \min f_1(x) &= \sum_{k=1}^{K} u_k \quad (\mathbf{DT}) \\ \min f_2(x) &= K \quad (\mathbf{DC}) \\ \min f_3(x) &= \sum_{k=1}^{K-1} \max_{1 \leq i \leq m} \max \left\{ |l_i^{k+1} - l_i^k|, |r_i^{k+1} - r_i^k| \right\} \quad (\mathbf{SU}_{var}) \\ \text{subject to } A &= \sum_{k=1}^{K} u_k S^k, u_k \in \mathbb{N}_0, S^k \in \mathcal{S}, \ \forall \, k \in [K] \end{cases}$$

We denote by $\mathcal{X}$ the feasible set in decision space and by $\mathcal{Y} = f(\mathcal{X})$ the feasible set in objective space.

In order to make the distinction between the different vectors of $\mathcal{Y}$, we use the dominance relation of Pareto. For two points $y^1, y^2$ in $\mathcal{Y}$, we say that $y^1$ dominates the point $y^2$ ($y^1 \prec y^2$) if and only if $y_k^1 \leq y_k^2 \ \forall k \in \{1, 2, 3\}$ and $y^1 \neq y^2$ (for at least one $k$, $y_k^1 < y_k^2$).

We can now introduce some definitions. A feasible solution $x^* \in \mathcal{X}$ is called *efficient* if there does not exist any other feasible solution $x \in \mathcal{X}$ such that $f(x) \prec f(x^*)$. The efficient set denoted by $X_E$ contains all the efficient solutions. The image $f(x^*)$ in objective space of an efficient solution $x^*$ is called a non-dominated point and the set of non-dominated points constitutes the Pareto front (or non-dominated frontier), denoted by $Y_N$. Supported efficient solutions are optimal solutions of the following weighted sum single-objective problem

$$\min_{x \in \mathcal{X}} \sum_{k=1}^{p} \lambda_k f_k(x)$$

for some vector $\lambda > 0$, that is each component is positive ($\lambda_k > 0, \forall k \in \{1, \ldots, p\}$). Non-supported efficient solutions are efficient solutions that are not optimal solutions of any weighted sum single-objective problem with $\lambda > 0$ (the images of these solutions are not located on the convex hull of the Pareto front). Therefore, these solutions are harder to compute than supported efficient ones.

Our aim is to generate a good approximation of the efficient set. If two solutions in the decision space give the same non-dominated point, only one solution is retained (only a good approximation of a minimal complete set [11] is generated).

To our knowledge, the multiobjective problem (P) has never been tackled before: nobody has tried to find the efficient solutions, or even a good approximation of the efficient solutions of (P).

Before presenting how we have solved this problem, it is interesting to show that the objectives can be conflicting. We have created the following example with this aim.

The matrix $A$ to decompose is the following:

$$A = \begin{pmatrix} 3 & 2 & 3 \\ 2 & 5 & 1 \end{pmatrix}$$

Let us consider the three following feasible decompositions:

- $D_1$:
$$1 \begin{pmatrix} 1 & 1 & 1 \\ 0 & 1 & 1 \end{pmatrix} + 1 \begin{pmatrix} 1 & 1 & 0 \\ 1 & 1 & 0 \end{pmatrix} + 1 \begin{pmatrix} 1 & 0 & 0 \\ 1 & 1 & 0 \end{pmatrix} + 2 \begin{pmatrix} 0 & 0 & 1 \\ 0 & 1 & 0 \end{pmatrix}$$

  with DT=5, DC=4 and $SU_{var}$=(1+1+2)=4.

- $D_2$:
$$2 \begin{pmatrix} 0 & 1 & 1 \\ 1 & 1 & 0 \end{pmatrix} + 3 \begin{pmatrix} 1 & 0 & 0 \\ 0 & 1 & 0 \end{pmatrix} + 1 \begin{pmatrix} 0 & 0 & 1 \\ 0 & 0 & 1 \end{pmatrix}$$

  with DT=6, DC=3 and $SU_{var}$=(2+2)=4.

- $D_3$:
$$3 \begin{pmatrix} 1 & 0 & 0 \\ 0 & 1 & 0 \end{pmatrix} + 2 \begin{pmatrix} 0 & 1 & 0 \\ 1 & 0 & 0 \end{pmatrix} + 2 \begin{pmatrix} 0 & 0 & 1 \\ 0 & 1 & 0 \end{pmatrix} + 1 \begin{pmatrix} 0 & 0 & 1 \\ 0 & 0 & 1 \end{pmatrix}$$

  with DT=8, DC=4 and $SU_{var}$=(1+1+1)=3.

**Property 1** *The optimal value of DT for the matrix $A$ is equal to 5.*

**Proof 1** See Engel's property (Property 7) in the next section.

The solution $D_1$ is thus an optimal solution for DT.

**Property 2** *The optimal value of DC for the matrix $A$ is equal to 3.*

**Proof 2** Suppose that a decomposition with two segments exists, with two different coefficients $a$ and $b$. By adding these two segments, we can generate a matrix with at most four different numbers equal to $(0, a, b, a+b)$. As the matrix $A$ contains four different numbers equal to $(1, 2, 3, 5)$ but not zero, it is impossible to find a decomposition with two segments.

The solution $D_2$ is thus an optimal solution for DC.

**Property 3** *The solution $D_2$ is the unique optimal solution for DC.*

**Proof 3** Because of the C1 constraint, the production of the two 3 of the matrix $A$ has to be realized with two different combinations of segments. Given that only three segments can be used, there are two possibilities: the first one is to use two segments with coefficients respectively equal to 2 and 1 and one segment with a coefficient equal to 3. This possibility is used in the solution $D_2$. The other possibility is to use one segment with a coefficient equal to 3 and another with a coefficient also equal to 3. In this case, the only possibility to produce the 5 of the matrix $A$ is to use a segment with a coefficient equal to 2 or 5. But that would make impossible the generation of the 1. Therefore, the decomposition $D_2$ is the only solution with DC equal to 3.

**Property 4** *For the matrix A, it is impossible to find a solution optimal for DT and DC at the same time.*

**Proof 4** Directly from Property 3, since the solution $D_2$ is the unique optimal solution for DC and this solution is not optimal for DT.

**Property 5** *The optimal value of $SU_{var}$ for the matrix A is equal to 3.*

**Proof 5** Suppose that a decomposition with $SU_{var}$ equal to 2 exists. This decomposition has to be composed of at most three segments. But as it is impossible to obtain a decomposition with less than three segments (see Property 2), this decomposition has to be composed of three segments. As the decomposition $D_2$ is the only solution with DC equal to 3 and that the $SU_{var}$ value of $D_2$ is equal to 4 and cannot be improved by changing the sequence (as the distance between each segment is always equal to 2), a decomposition with $SU_{var}$ equal to 2 does not exist. $SU_{var}$ equal to 3 is thus the optimal value.

The solution $D_3$ is thus an optimal solution for $SU_{var}$.

We can wonder whether a solution optimal for DT and $SU_{var}$ at the same time exists, that is with a DT value equal to 5 and a $SU_{var}$ value equal to 3.

**Property 6** *For the matrix A, it is impossible to find a solution optimal for DT and $SU_{var}$ at the same time.*

**Proof 6** As consequence of Properties 3 and 4, an optimal solution for DT cannot be composed of three segments. Therefore, a decomposition optimal for DT and $SU_{var}$ should be composed of four segments (if the number of segments is higher than four, the $SU_{var}$ value would be necessarily superior to 3). But the only four coefficients $a, b, c$ and $d$ ($\in \mathbb{N}_0$) that respect $(a + b + c + d = 5)$ are $(2, 1, 1, 1)$, that is the coefficients of the decomposition $D_1$. As $D_1$ holds a $\text{SU}_{var}$ value equal to 4 that cannot be improved (the distance between the last segment and the other segments is always equal to 2), it is impossible to find a solution which is optimal for DT and $\text{SU}_{var}$ at the same time for the matrix $A$.

We have thus shown, for this example, that one solution optimizing DT has not an optimal value for DC or $\text{SU}_{var}$ and that the unique solution optimizing DC has not an optimal value for DT or $\text{SU}_{var}$. And therefore, a solution that optimizes $\text{SU}_{var}$ has not an optimal value for DT or DC. The objectives of this example are thus conflicting. As far as we know, this is the smallest example in size (3x2) and also for the maximal value of the matrix (5) that holds this property.

The three non-dominated points of this example are (5,4,4), (6,3,4) and (8,4,3) (it is easy to check that we cannot find a non-dominated point with DT=7).

## 3  Two-phase Pareto local search

To solve the MOCO problem (P), we use the two-phase Pareto local search (2PPLS) that we have recently developed [18]. This method has enabled to obtain state-of-the-art results for the multiobjective traveling salesman problem [16, 18] and the multiobjective multidimensional knapsack problem [17].

In the first phase of the method, we generate a good approximation of the supported efficient solutions, by using weighted sums and efficient single-objective solvers for optimizing the weighted single-objective problems. The aim of the second phase is to generate non-supported efficient solutions. With this aim, we use the Pareto local search (PLS) [1, 20]. This method is a straightforward adaptation of local search to the multiobjective case and only needs a neighborhood. At the end, a local optimum, defined in a multiobjective context, is found [20] (called a Pareto local optimum set). The method does not require any objectives aggregation nor any numerical parameters. By applying PLS to connect the supported efficient solutions previously produced, non-supported efficient solutions can be generated. Indeed, efficient solutions of MOCO problems are often located in the same search space and a neighborhood integrated in PLS allows to explore efficiently this search space. The main part of the adaptation of 2PPLS to MOCO problems concerns the definition of the neighborhood, that needs a good compromise between running time and quality of the results obtained.

The pseudo-code of 2PPLS is given by the algorithm 1.

**Algorithm 1** 2PPLS

    Parameters ↓: an initial population $P_0$, a neighborhood function $\mathcal{N}(x)$.
    Parameters ↑: an approximation $\widetilde{X}_E$ of the efficient set $X_E$.

    --| Initialization of $\widetilde{X}_E$ and a population $P$ with the initial population $P_0$
    $\widetilde{X}_E \leftarrow P_0$
    $P \leftarrow P_0$
    --| Initialization of an auxiliary population $P_a$
    $P_a \leftarrow \varnothing$
    **while** $P \neq \varnothing$ **do**
        --| Generation of all neighbors $p'$ of each solution $p \in P$
        **for all** $p \in P$ **do**
            **for all** $p' \in \mathcal{N}(p)$ **do**
                **if** $f(p) \not\leq f(p')$ **then**
                      `AddSolution`($\widetilde{X}_E$ ↕, $p'$ ↓, $f(p')$ ↓, $Added$ ↑)
                      **if** $Added = true$ **then**
                            `AddSolution`($P_a$ ↕, $p'$ ↓, $f(p')$ ↓)
        --| $P$ is composed of the new potentially efficient solutions
        $P \leftarrow P_a$
        --| Reinitialization of $P_a$
        $P_a \leftarrow \varnothing$

The method starts with the population $P$ composed of potentially efficient solutions given by the initial population $P_0$. Then, all the neighbors $p'$ of each solution $p$ of $P$ are generated. If a neighbor $p'$ is not weakly dominated by the current solution $p$, we try to add the solution $p'$ to the approximation $\widetilde{X}_E$ of the efficient set, which is updated with the procedure `AddSolution`. This procedure is not described in this paper but simply consists of updating an approximation $\widetilde{X}_E$ of the efficient set when a new solution $p$ is added to $\widetilde{X}_E$. This procedure has four parameters: the set $\widetilde{X}_E$ to actualize, the new solution $p$, its evaluation $f(p)$ and a boolean variable called $Added$ that returns $True$ if the new solution has been added and $False$ otherwise. If the solution $p'$ has been added to $\widetilde{X}_E$, the boolean variable $Added$ is true and the solution $p'$ is added to $P_a$, which is also updated with the procedure `AddSolution`. Therefore, $P_a$ is only composed of (new) potentially efficient solutions. Once all the neighbors of each solution of $P$ have been generated, the algorithm starts again, with $P$ equal to $P_a$, until $P = P_a = \varnothing$. The auxiliary population $P_a$ is used such that the neighborhood of each solution of the population $P$ is explored, even if some solutions of $P$ become dominated following the addition of a new solution to $P_a$. Thus, sometimes, neighbors are generated from a dominated solution.

# 4 Adaptation of 2PPLS to the multiobjective decomposition problem

In this section, we present how 2PPLS has been adapted to the problem (P). We first present how the initial population has been generated. We expose then the neighborhood needed in PLS, method used as second phase in 2PPLS.

## 4.1 Initial population

As no efficient method for solving the single-objective problem resulting from the linear aggregation of the three objectives considered (DT,DC,SU$_{var}$) is known, we use another technique to generate the initial population.

Two initial solutions of good quality are generated and used to form the initial population. The first solution is a good approximation of lexmin(DT,DC,SU$_{var}$) and the second one is a good approximation of lexmin(DT,SU$_{var}$,DC). In both cases, we first minimize DT since polynomial algorithms are known for this problem.

### 4.1.1 Approximation of lexmin(DT,DC,SU$_{var}$)

To approximate lexmin(DT,DC,SU$_{var}$), we first approximate lexmin(DT,DC) with the heuristic of Engel [9] which is efficient for this problem. We then apply a TSP heuristic to reduce the SU$_{var}$ value of the solution. We use a very efficient heuristic for the TSP: the Lin and Kernighan heuristic implemented by Helsgaun (LKH) [13].

The algorithm developed by Engel is a deterministic construction algorithm, that allows to find an optimal solution for the DT objective with a low DC value. Engel tackles this problem as follows: he removes different well-selected combinations of couples $(u_t, S^t)$ from the current matrix until $A_{t+1} = 0$, with $A_{t+1} = A_t - u_t S^t$, where $t$ represents the index of the step of the construction algorithm. He starts the algorithm with $A_0 = A$. A move consists thus of removing from the current matrix a segment multiplied by a certain coefficient.

At each step of the construction heuristic, the maximum coefficient ($u_{max}$) that can be used while ensuring that the optimal objective DT can be achieved is considered. Using $u_{max}$ allows to find very good results for the lexicographic problem (DT,DC) [8, 9, 14], since using decompositions with high coefficient values increases the chances to find a solution with a low value of DC. Unfortunately, no good quality upper bound for the DC values obtained with this heuristic has been found.

Engel has developed a theory to compute $u_{max}$. The coefficient $u_{max}$ can be easily obtained in $O(mn^2)$.

We present below the basic results of Engel on which are based the computation of the coefficient $u_{max}$ (we do not go into details).

If two zero columns are added to $A$, that is let:

$$a_{i,0} = a_{i,n+1} = 0 \quad \forall\, i \in [m]$$

we can associate to $A$ its difference matrix $D$ of dimension $m \times (n+1)$:

$$d_{i,j} = a_{i,j} - a_{i,j-1} \quad \forall\, i \in [m],\ \forall\, j \in [n+1]$$

The row complexity $c_i(A)$ of $A$ is defined by

$$c_i(A) = \sum_{j=1}^{n} \max\{0, d_{i,j}\}$$

and the complexity $c(A)$ of $A$ is the maximum of the row complexity:

$$c(A) = \max_{i \in [m]} c_i(A)$$

**Property 7** *(Engel's property)* *The complexity $c(A)$ is equal to the optimal DT value that we can obtain for a decomposition of $A$ [9].*

For instance, for the matrix of the preceding section:

$$\begin{pmatrix} 3 & 2 & 3 \\ 2 & 5 & 1 \end{pmatrix}$$

the difference matrix is equal to:

$$D = \begin{pmatrix} 3 & -1 & 1 & -3 \\ 2 & 3 & -4 & -1 \end{pmatrix}$$

The complexity $c_1$ of the first row is equal to 4 while the complexity $c_2$ of the second row is equal to 5. Therefore, the complexity of the matrix is equal to 5 ($= \max(4,5)$), which is equal to the optimal value for DT.

In the heuristic of Engel, to keep the optimal DT value, the coefficient $u_{max}$ has to respect the following properties:

$$c(A - u_{max}S) = c(A) - u_{max}, \quad \text{that is}$$

$$c_i(A - u_{max}S) \leq c(A) - u_{max} \quad \forall\, i \in [m]$$

Engel gives a $O(mn^2)$ algorithm to compute $u_{max}$ based on the preceding relations (the algorithm is not detailed here).

But once $u_{max}$ has been defined, we also have to define which segments to use among all segments that respect $u_{max}$. Kalinoswki [14] has developed a trial and error rule that gives slightly better results that the initial rule of Engel. The rule is as follows.

He computes the scalar $q$ associated to $A$ as following:

$$q(A) = \left|\{(i,j) \in [m] \times [n] : d_{i,j} \neq 0\}\right|,$$

and in his method, a segment $S$ that minimizes $q(A - u_{max}S)$ is selected.

This method gives very good results for lexmin(DT,DC), but these results are only justified by experiments.

*4.1.2 Approximation of lexmin(DT,$SU_{var}$,DC)*

To approximate lexmin(DT,$SU_{var}$,DC), we propose to adapt the heuristic of Engel since $SU_{var}$ is correlated to the DC objective. We keep the principle of the construction algorithm of Engel by removing well-selected combinations of couples $(u_t, S^t)$ from the current matrix until $A_{t+1} = 0$. As in the Engel algorithm, it is worthwhile to take the maximal coefficient $u_{max}$ that allows to keep the minimal DT value. For the definition of the segment that corresponds to $u_{max}$, we try three new rules, presented below.

For each line of the new segment to define, we have to choose between different $(l_i, r_i)$ intervals. We sort the intervals as follows: $\{(0,1), (0,2), \cdots, (0, n+1), (1,3), \cdots, (1, n+1), \cdots, (n-2, n), (n-2, n+1), (n-1, n+1)\}$ (see Fig. 2 for an example for a line of length 4) and we define three rules from this sorting:

- The first rule is to take the first feasible interval (that is an interval that does not irradiate too much an entry of the intensity matrix to decompose), following this order.
- The second rule is to take the last feasible interval, following this order.
- The third rule is to take the first feasible interval which is the closest to the preceding interval of the same line (that is minimizing $\max\{|l_i^{k+1} - l_i^k|, |r_i^{k+1} - r_i^k|\}$): the aim is to minimize the maximal distance between two consecutive segments. The first interval defined with this rule is the same as the one selected with the first rule.

For the first two rules, we expect to always move the leaves from the left to the right or to the right from the left in order to minimize the maximal distance

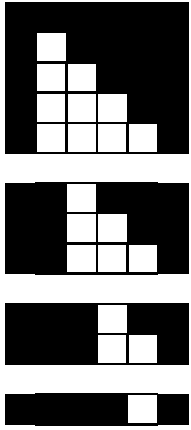

Fig. 2. Specific order of the leaves for a line of length 4.

between two consecutive segments.

The results of the comparison between the different rules will be given in section 5.

*4.2 Neighborhood*

The neighborhood is the main component of PLS (employed in the second phase of 2PPLS). It is not trivial to produce neighbors from a feasible solution for the problem (P) considered. Indeed, removing one segment from the current solution is not enough to produce a solution different than the current solution. Removing two segments from the current solution requires to determine how to select both segments and how to recombine these two segments to produce a new feasible solution, ideally of better quality. Moreover, it will be difficult with this kind of technique to find neighbors with different DC values. We can try to modify the sequence of segments in order to improve the $SU_{var}$ objective, but it is always possible to apply a TSP algorithm at the end of the decomposition to improve this objective.

The neighborhood developed in this work is thus a bit complex. It works as follows, for the generation of one neighbor from a current decomposition:

(1) Selection of a segment $S$ that belongs to the current decomposition.
(2) We modify a line $i$ of $S$ in the following way:

$$l_i = l_i + (-1 \text{ or } 0 \text{ or } 1)$$
$$r_i = r_i + (-1 \text{ or } 0 \text{ or } 1)$$

(3) We put $S$ at the first place of the new decomposition.
(4) We eventually modify the coefficient of this segment.
(5) We construct a neighbor by adding the segments in the order of the

current decomposition. If a segment is not feasible, we skip it. We adapt the coefficient of the segments added, which is equal to the minimum between the current coefficient of the segment and the maximal feasible coefficient. This rule has been adopted since the current coefficient comes from a preceding solution of good quality.
(6) The matrix that remains after adding all possible segments is decomposed with the heuristic of Engel.
(7) Once a decomposition is obtained, we optimize the $SU_{var}$ objective by using a simple and fast TSP heuristic: the first improvement local search based on the two-edge exchange moves [12].

This neighborhood requires the definition of many elements, but we will see that it is possible to explore many possibilities, in order to produce many neighbors.

To illustrate the neighborhood, we show its functioning on the following example:

$$A = \begin{pmatrix} 8 & 5 & 6 \\ 5 & 3 & 6 \end{pmatrix}$$

The difference matrix of $A$ is equal to:

$$D = \begin{pmatrix} 8 & -3 & 1 & -6 \\ 5 & -2 & 3 & -6 \end{pmatrix}$$

The complexity $c_1$ of the first row is equal to 9 while the complexity $c_2$ of the second row is equal to 8. The complexity of the matrix is thus equal to 9 (=max(8, 9)), which is equal to the optimal value for DT.

Let us consider that we start the neighborhood from one of the solutions that minimize the DT objective:

$$3 \begin{pmatrix} 1 & 0 & 0 \\ 0 & 0 & 1 \end{pmatrix} + 1 \begin{pmatrix} 0 & 0 & 1 \\ 0 & 1 & 1 \end{pmatrix} + 3 \begin{pmatrix} 1 & 1 & 1 \\ 1 & 0 & 0 \end{pmatrix} + 2 \begin{pmatrix} 1 & 1 & 1 \\ 1 & 1 & 1 \end{pmatrix}$$

The DT value of this solution is equal to 9. DC is equal to 4 and $SU_{var}$ is equal to (2+2+2)=6.

We apply the neighborhood to this solution:

(1) We select the first segment:
$$\begin{pmatrix} 1 & 0 & 0 \\ 0 & 0 & 1 \end{pmatrix}$$

(2) We select the second line. For this line, $l = 2$ and $r = 4$. We modify this line by putting $l = 1$, $r = 3$. We obtain this segment:
$$\begin{pmatrix} 1 & 0 & 0 \\ 0 & 1 & 0 \end{pmatrix}$$

(3) We put this segment at the first place of the new decomposition.
(4) The current coefficient of this segment is equal to 3. As for this segment, the maximal feasible coefficient that we can put is 3, we keep the coefficient equal to 3. The remaining matrix is:
$$\begin{pmatrix} 5 & 5 & 6 \\ 5 & 0 & 6 \end{pmatrix}$$

(5) The first segment that we can consider is:
$$\begin{pmatrix} 0 & 0 & 1 \\ 0 & 1 & 1 \end{pmatrix}$$
but we cannot add it.

The second segment is:
$$\begin{pmatrix} 1 & 1 & 1 \\ 1 & 0 & 0 \end{pmatrix}$$
We add it, with a coefficient equal to 3, that is its current coefficient (the maximal feasible coefficient is equal to 5). The remaining matrix is:
$$\begin{pmatrix} 2 & 2 & 3 \\ 2 & 0 & 6 \end{pmatrix}$$

The last segment in the initial decomposition is:
$$\begin{pmatrix} 1 & 1 & 1 \\ 1 & 1 & 1 \end{pmatrix}$$
but we cannot add it.

(6) The decomposition of the remaining matrix with the heuristic of Engel gives:

$$5\begin{pmatrix} 0\ 0\ 0 \\ 0\ 0\ 1 \end{pmatrix} + 2\begin{pmatrix} 1\ 1\ 1 \\ 1\ 0\ 0 \end{pmatrix} + 1\begin{pmatrix} 0\ 0\ 1 \\ 0\ 0\ 1 \end{pmatrix}$$

The decomposition obtained is thus:

$$3\begin{pmatrix} 1\ 0\ 0 \\ 0\ 1\ 0 \end{pmatrix} + 3\begin{pmatrix} 1\ 1\ 1 \\ 1\ 0\ 0 \end{pmatrix} + 5\begin{pmatrix} 0\ 0\ 0 \\ 0\ 0\ 1 \end{pmatrix} + 2\begin{pmatrix} 1\ 1\ 1 \\ 1\ 0\ 0 \end{pmatrix} + 1\begin{pmatrix} 0\ 0\ 1 \\ 0\ 0\ 1 \end{pmatrix}$$

But as we can see, it is possible to combine the second segment with the fourth one, to obtain this decomposition:

$$3\begin{pmatrix} 1\ 0\ 0 \\ 0\ 1\ 0 \end{pmatrix} + 5\begin{pmatrix} 1\ 1\ 1 \\ 1\ 0\ 0 \end{pmatrix} + 5\begin{pmatrix} 0\ 0\ 0 \\ 0\ 0\ 1 \end{pmatrix} + 1\begin{pmatrix} 0\ 0\ 1 \\ 0\ 0\ 1 \end{pmatrix}$$

Therefore, each time we try to add a segment to a current decomposition, we check if we can combine this segment with other segments of the decomposition, in order to reduce the DC and $SU_{var}$ values of the decomposition.

The DT value of this solution is equal to 14, DC is equal to 4 and $SU_{var}$ is equal to (2+3+3)=8.

(7) The matrix of distances between the segments is as follows:

$$\begin{bmatrix} & S_1 & S_2 & S_3 & S_4 \\ S_1 & 0 & 2 & 1 & 2 \\ S_2 & 2 & 0 & 3 & 2 \\ S_3 & 1 & 3 & 0 & 3 \\ S_4 & 2 & 2 & 3 & 0 \end{bmatrix}$$

We see that by doing a two-edge exchange move, we can obtain the new sequence $(S_3, S_1, S_2, S_4)$, that improves the $SU_{var}$ objective by 3 units. We obtain the following neighbor:

$$5\begin{pmatrix} 0\ 0\ 0 \\ 0\ 0\ 1 \end{pmatrix} + 3\begin{pmatrix} 1\ 0\ 0 \\ 0\ 1\ 0 \end{pmatrix} + 5\begin{pmatrix} 1\ 1\ 1 \\ 1\ 0\ 0 \end{pmatrix} + 1\begin{pmatrix} 0\ 0\ 1 \\ 0\ 0\ 1 \end{pmatrix}$$

The evaluation vector of this solution is thus equal to (14,4,5).

We have therefore obtained a new potentially efficient solution, and we can again apply the neighborhood from this solution:

(1) We select the third segment:

$$\begin{pmatrix} 1\ 1\ 1 \\ 1\ 0\ 0 \end{pmatrix}$$

(2) We select the first line. For this line, $l = 0$ and $r = 4$. We modify this line by putting $r = 3$. We obtain this segment:

$$\begin{pmatrix} 1\ 1\ 0 \\ 1\ 0\ 0 \end{pmatrix}$$

(3) We put this segment at the first place of the new decomposition.
(4) The current coefficient of this segment is equal to 5. As for this segment, the maximal feasible coefficient that we can put is 5, we keep the coefficient equal to 5. The remaining matrix is:

$$\begin{pmatrix} 3\ 0\ 6 \\ 0\ 3\ 6 \end{pmatrix}$$

(5) The first segment that we can consider is:

$$\begin{pmatrix} 0\ 0\ 0 \\ 0\ 0\ 1 \end{pmatrix}$$

We add it, with a coefficient equal to 5, that is its current coefficient (the maximal feasible coefficient is equal to 6). The remaining matrix is:

$$\begin{pmatrix} 3\ 0\ 6 \\ 0\ 3\ 1 \end{pmatrix}$$

The second segment is:

$$\begin{pmatrix} 1\ 0\ 0 \\ 0\ 1\ 0 \end{pmatrix}$$

We add it, with a coefficient equal to 3, that is its current coefficient (the maximal feasible coefficient is equal to 3). The remaining matrix is:

$$\begin{pmatrix} 0\ 0\ 6 \\ 0\ 0\ 1 \end{pmatrix}$$

The last segment in the initial decomposition is:

$$\begin{pmatrix} 0\ 0\ 1 \\ 0\ 0\ 1 \end{pmatrix}$$

We add it, with a coefficient equal to 1, that is its current coefficient (the maximal feasible coefficient is equal to 1).

(6) The remaining matrix is:

$$\begin{pmatrix} 0\ 0\ 5 \\ 0\ 0\ 0 \end{pmatrix} = 5 \begin{pmatrix} 0\ 0\ 1 \\ 0\ 0\ 0 \end{pmatrix}$$

The decomposition obtained is thus:

$$5 \begin{pmatrix} 1\ 1\ 0 \\ 1\ 0\ 0 \end{pmatrix} + 5 \begin{pmatrix} 0\ 0\ 0 \\ 0\ 0\ 1 \end{pmatrix} + 3 \begin{pmatrix} 1\ 0\ 0 \\ 0\ 1\ 0 \end{pmatrix} + 1 \begin{pmatrix} 0\ 0\ 1 \\ 0\ 0\ 1 \end{pmatrix} + 5 \begin{pmatrix} 0\ 0\ 1 \\ 0\ 0\ 0 \end{pmatrix}$$

But as we can see, it is possible to combine the second segment with the last one, and then the result of this combination with the fourth segment to obtain this decomposition:

$$5 \begin{pmatrix} 1\ 1\ 0 \\ 1\ 0\ 0 \end{pmatrix} + 3 \begin{pmatrix} 1\ 0\ 0 \\ 0\ 1\ 0 \end{pmatrix} + 6 \begin{pmatrix} 0\ 0\ 1 \\ 0\ 0\ 1 \end{pmatrix}$$

with a $SU_{var}$ value equal to (1+2)=3.

(7) The matrix of distances between the segments is as follows:

$$\begin{bmatrix} & S_1 & S_2 & S_3 \\ S_1 & 0 & 1 & 2 \\ S_2 & 1 & 0 & 2 \\ S_3 & 2 & 2 & 0 \end{bmatrix}$$

We see that it is impossible to improve the $SU_{var}$ objective by doing two-edge exchange moves.

The DT value of this solution is equal to 14, and the DC and $SU_{var}$ values are equal to 3 (which are the optimal values of these two objectives [8]).

Therefore, by applying two times the neighborhood from a solution that minimizes the DT objective, we have found a solution that minimizes the DC and $SU_{var}$ values.

It should be noted that we have found one more non-dominated point for this problem: the point (10,4,4), that can be obtained with this solution:

$$2 \begin{pmatrix} 0\ 0\ 1 \\ 1\ 1\ 1 \end{pmatrix} + 3 \begin{pmatrix} 1\ 0\ 0 \\ 1\ 0\ 0 \end{pmatrix} + 1 \begin{pmatrix} 1\ 1\ 0 \\ 0\ 1\ 0 \end{pmatrix} + 4 \begin{pmatrix} 1\ 1\ 1 \\ 0\ 0\ 1 \end{pmatrix}$$

For this small problem, we have therefore found three non-dominated points: (9,4,6), (10,4,4) and (14,3,3).

### 4.3 Final optimization step

For each potentially efficient solution found at the end of 2PPLS, we apply the LKH heuristic, to eventually improve the $\mathrm{SU}_{var}$ value of the solutions.

## 5 Results

### 5.1 Instances

We use two types of instances: random instances and real instances. The random instances are the same instances that Engel and Kalinowski [9, 14] used to test their algorithms for lexmin(DT,DC). That enables us to check that our implementation of the heuristic of Engel gives the same results. The random instances are matrices 15x15 with randomly generated elements (uniformly distributed) between zero and the parameter $L$ ($L$ varies from 3 to 16 as done by Engel and Kalinowski).

The real instances come from the radiology department of the "Klinikum für Strahlentherapie" in Rostock[2]. The instances correspond to radiotherapy plans for the treatment of prostate cancers. We use 25 different instances whose the size varies from 12x21 to 23x20, and whose the maximal value varies from 13 to 31 (the minimal value is always equal to 0).

We first experiment the different rules for the adaptation of the heuristic of Engel (see section 4.1.2). We then expose the results obtained with 2PPLS.

---

[2] We thank Antje Keisel from the Institue for Mathematics of the University of Rostock to have provided us these instances.

## 5.2 Rules for the heuristic of Engel

We experiment here the different rules for the selection of the intervals in the heuristic of Engel. The rule "Min" is to take the first feasible interval which is the closest to the preceding interval. The rule "First"("Last") is to take the first (last) feasible interval according to the order defined in section 4.1.2. The rule "Kali" is the rule developed by Kalinowski.

### 5.2.1 Instances of Engel

To compare the four rules, we first use the instances of Engel. For each value of $L$ we make the average on 1000 different matrices for three values: the DC objective, the $SU_{var}$ objective and the $SU_{var}$ optimized objective which is the value of $SU_{var}$ obtained after optimization of the sequence of segments with the LKH heuristic. For each matrix, as the heuristic is mainly deterministic (only the final optimization of $SU_{var}$ with LKH is not), only one execution is run.

The results are given in Table 1 for DC and in Table 2 for $SU_{var}$ and $SU_{var}$ optimized.

Table 1
Average values of DC obtained by the different rules for the random instances ($A$=15x15).

| $L$ | DC | | | |
|---|---|---|---|---|
| | Min | First | Last | Kali |
| 3 | 10.32 | 10.32 | 9.93 | **9.72** |
| 4 | 11.73 | 11.74 | 11.33 | **10.94** |
| 5 | 12.69 | 12.69 | 12.29 | **11.76** |
| 6 | 13.56 | 13.56 | 13.14 | **12.50** |
| 7 | 14.27 | 14.26 | 13.85 | **13.12** |
| 8 | 14.91 | 14.90 | 14.48 | **13.71** |
| 9 | 15.52 | 15.46 | 15.09 | **14.20** |
| 10 | 16.04 | 15.98 | 15.59 | **14.69** |
| 11 | 16.46 | 16.37 | 16.01 | **15.06** |
| 12 | 16.92 | 16.81 | 16.49 | **15.46** |
| 13 | 17.30 | 17.18 | 16.87 | **15.81** |
| 14 | 17.62 | 17.53 | 17.21 | **16.13** |
| 15 | 18.01 | 17.91 | 17.55 | **16.52** |
| 16 | 18.29 | 18.16 | 17.89 | **16.79** |

We remark that for the DC objective, the "Kali" rule is better than the others. That is logical since the "Kali" rule is suited to minimize DC. And we have, except for $L = 4$, Kali $\succ$ Last $\succ$ First $\succeq$ Min.

Table 2
Average values of SU$_{var}$ and SU$_{var}$ optimized obtained by the different rules for the random instances ($A$=15x15).

|   | SU$_{var}$ | | | | SU$_{var}$ optimized | | | |
|---|---|---|---|---|---|---|---|---|
| $L$ | Min | First | Last | Kali | Min | First | Last | Kali |
| 3 | 79.69 | **69.57** | 70.38 | 113.68 | 77.43 | 63.05 | **61.23** | 103.14 |
| 4 | 94.22 | **83.74** | 84.10 | 127.63 | 91.72 | 74.72 | **71.63** | 114.44 |
| 5 | 104.81 | 96.55 | **95.11** | 137.43 | 101.82 | 84.85 | **80.34** | 122.28 |
| 6 | 112.35 | 107.29 | **105.53** | 146.62 | 108.98 | 93.19 | **88.62** | 129.35 |
| 7 | 119.49 | 117.02 | **114.22** | 153.55 | 115.75 | 101.14 | **95.83** | 134.90 |
| 8 | 126.53 | 126.12 | **122.94** | 160.69 | 122.39 | 108.12 | **102.30** | 140.40 |
| 9 | 133.32 | 133.65 | **129.75** | 166.67 | 128.38 | 114.32 | **107.66** | 144.95 |
| 10 | 139.10 | 141.08 | **137.05** | 172.46 | 134.04 | 120.17 | **113.93** | 149.60 |
| 11 | 144.16 | 147.09 | **142.57** | 177.08 | 138.56 | 125.22 | **118.31** | 152.92 |
| 12 | 149.41 | 153.17 | **148.31** | 182.08 | 143.47 | 130.06 | **122.71** | 156.71 |
| 13 | 153.56 | 158.54 | **153.42** | 186.16 | 147.34 | 134.47 | **126.85** | 160.10 |
| 14 | **157.19** | 162.91 | 157.96 | 189.64 | 150.70 | 138.10 | **130.82** | 162.75 |
| 15 | **162.00** | 168.27 | 163.02 | 194.47 | 154.91 | 142.46 | **135.04** | 166.35 |
| 16 | **166.22** | 172.34 | 167.41 | 197.64 | 158.84 | 146.25 | **138.53** | 169.18 |

On the other hand, if we want to minimize SU$_{var}$, we remark that the "Last" rule allows to obtain better results for the SU$_{var}$ optimized objective. And we have Last ≻ First ≻ Min ≻ Kali.

If we do not consider the optimization step, the rule "First" is the best for $L = 3$ and $L = 4$, the "Last" rule is the best for $L$ going from 5 to 13 and the rule "Min" is the best for $L$ going from 14 to 16. The "Kali" rule is always the worst.

The "Last" rule seems thus better than the rule of Kalinowski for the SU$_{var}$ objective, even if this rule gives higher DC value on average. The running time of the heuristic of Engel with each of these rules is negligible.

In Fig. 3, we show the evolution of SU$_{var}$ optimized (average on 100 instances) according to $L$, going here from 10 to 10000 (in order to study the asymptotic behavior of the rules), for a 15x15 matrix. The "Last" and "Kali" rules are compared. We see on this figure that the difference between the "Last" and "Kali" rules is not so well marked than in Table 2. When $L$ is bigger, there is anymore strong difference between both rules and in some cases, the "Kali" rule gives better results for the SU$_{var}$ optimized objective.

We show in Fig. 4 the evolution of SU$_{var}$ optimized (average on 100 instances) according to the size of the matrix $A$. The value of $L$ is fixed to 10. The matrix $A$ is always considered as square and its size varies from 3x3 to 50x50. We see on this figure that here, the difference between the "Last" and "Kali" rules is more marked when the size of the matrix increases. The results given by the "Kali" rule are getting worse in comparison with the "Last" rule.

### 5.2.2 Real instances

For the 25 real instances, the results are not reported here but we have found that:

- For the DC value, the "Kali" rule that gives the best results except on five instances for which the "Last" rule gives a DC value lower of one or two units.

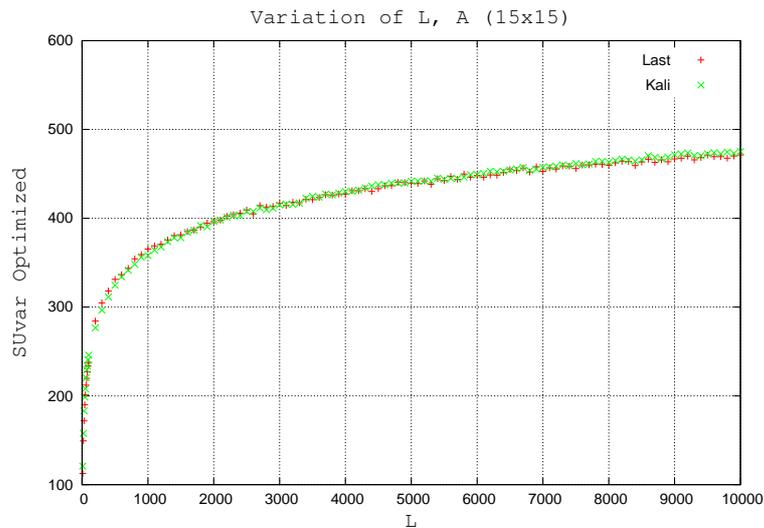

Fig. 3. Evolution of $SU_{var}$ according to $L$ for a 15x15 matrix. The "Last" and "Kali" rules are compared.

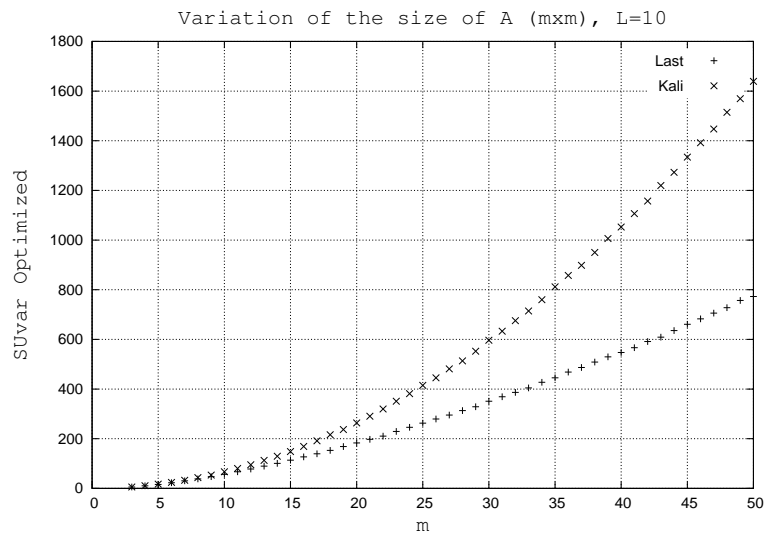

Fig. 4. Evolution of $SU_{var}$ according to the size of the square matrix $m$ x $m$. The "Last" and "Kali" rules are compared.

- For SU$_{var}$ without optimization, on 19 instances the "Min" rule gives the best results. With optimization, the "Last" rule gives better results on 17 instances. The "Kali" rule never gives the best results.

### 5.3 Two-phase Pareto local search

We experiment in this section 2PPLS.

The initial population of 2PPLS is formed with two solutions (see section 4.1). If there is one solution that dominates the other, the initial population will be composed of only one solution: the non-dominated one.

For the neighborhood, we adopt the following choices:

- We try all possible segments for the segment that we put at the beginning of the new decomposition.
- Either we do not modify the segment or we modify it by trying all possibilities of modification. We modify each line of the segment separately, by considering all feasible possibilities for each line (equal to maximum 8): $\big((+1,+1),(+1,0),(+1,-1),(0,+1),(0,-1),(-1,+1),(-1,0),(-1,-1)\big)$.
- We try all feasible coefficients for the segment that we put at the beginning of the decomposition.
- The remaining matrix is decomposed with the heuristic of Engel, with the "Last" rule.

If the number of segments of the current decomposition is equal to $K$ and the maximal value of the matrix equal to $L$, a crude bound for the number of neighbors generated is thus equal to $KL(8m+1)$.

As we explore entirely the neighborhood, 2PPLS is mainly deterministic (only the final optimization step is not) and will be applied only one time for each intensity matrix.

#### 5.3.1 Instances of Engel

We first experiment the method on the random instances of Engel. As no state-of-the-art results are known for this multiobjective problem, we use very specific indicators to measure the quality of the approximations obtained.

In Table 3, we report the number of potentially efficient solutions found ($|PE|$), the number of phases of PLS before convergence and the running time in seconds (on a Intel Core 2 Duo T7500 2.2 GHz CPUs and 2 GB of RAM). For each value of $L$, we indicate for each indicator the minimum, maximum and mean values found on 20 different matrices.

Table 3
Average values of the indicators obtained by 2PPLS for the random instances (1).

|   | $|PE|$ | | | Number of phases | | | Time(s) | | |
|---|---|---|---|---|---|---|---|---|---|
| $L$ | Mean | Min | Max | Mean | Min | Max | Mean | Min | Max |
| 3 | 1.35 | 1 | 2 | 3.60 | 1 | 7 | 1.34 | 0.26 | 3.29 |
| 4 | 1.80 | 1 | 3 | 5.45 | 1 | 10 | 3.48 | 0.42 | 8.14 |
| 5 | 1.85 | 1 | 3 | 5.30 | 1 | 11 | 5.08 | 0.57 | 12.94 |
| 6 | 2.00 | 1 | 5 | 6.25 | 2 | 16 | 7.17 | 1.63 | 16.34 |
| 7 | 2.40 | 1 | 5 | 5.60 | 1 | 10 | 8.70 | 1.19 | 28.55 |
| 8 | 2.30 | 1 | 4 | 5.65 | 1 | 10 | 11.12 | 1.20 | 24.94 |
| 9 | 2.75 | 1 | 7 | 5.40 | 1 | 10 | 12.82 | 1.28 | 35.83 |
| 10 | 2.55 | 1 | 7 | 5.40 | 2 | 10 | 15.52 | 2.82 | 44.85 |
| 11 | 2.70 | 1 | 6 | 6.35 | 1 | 12 | 19.52 | 1.69 | 49.66 |
| 12 | 2.40 | 1 | 5 | 4.55 | 1 | 12 | 17.37 | 1.86 | 61.76 |
| 13 | 2.80 | 1 | 6 | 6.95 | 1 | 13 | 25.31 | 2.25 | 53.04 |
| 14 | 3.30 | 2 | 5 | 6.20 | 3 | 12 | 28.12 | 11.61 | 59.17 |
| 15 | 3.15 | 2 | 7 | 6.30 | 3 | 10 | 34.30 | 8.95 | 123.83 |
| 16 | 3.20 | 1 | 7 | 6.40 | 1 | 16 | 32.77 | 2.82 | 75.59 |

We remark that:

- The number of potentially efficient solutions is not high: between 1 and 7, with a mean value between 1 and 4. The correlation between the objectives seems thus high for these instances.
- The number of phases is included between 1 and 16, with a mean value between 3 and 7, which means that the neighborhood is efficient since at each phase improvements are realized. Please remind that if there is no new potentially efficient solution generated during a phase, PLS stops since a Pareto local optimum set has been found. Moreover, we start PLS from relatively good initial solutions.
- The mean running time is acceptable, between 1 and 35 seconds. However, for some instances, the running time can be higher, until 124 seconds for an instance with $L = 15$.

The evaluation of the quality of the results is given in Table 4. We evaluate the improvements of DC by comparing the best values obtained with 2PPLS to the values obtained by the heuristic of Engel with the "Kali" rule (column "% DC - Kali"). We also evaluate the improvements of $SU_{var}$ by comparing the best $SU_{var}$ values obtained with 2PPLS to the $SU_{var}$ values obtained with the heuristic of Engel with the "Kali" rule and LKH (column "% $SU_{var}$ - Kali+LKH") and to the $SU_{var}$ values obtained with the heuristic of Engel with the "Last" rule and LKH (column "% $SU_{var}$ - Last+LKH"). Two cases are distinguished: initially we evaluate the improvements made if we keep the optimal value for DT (column "DT optimal"), and secondly, we have no restriction on the value of the DT objective (column "DT not necessary optimal").

Table 4
Average values of the indicators obtained by 2PPLS for the random instances (2).

|   | DT optimal | | | DT not necessary optimal | | |
| --- | --- | --- | --- | --- | --- | --- |
|   | % DC | % $SU_{var}$ | | % DC | % $SU_{var}$ | |
| $L$ | Kali | Kali+LKH | Last+LKH | Kali | Kali+LKH | Last+LKH |
| 3  | 0.00 | 42.97 | 4.83 | 0.45 | 43.22 | 5.32 |
| 4  | 0.42 | 38.74 | 4.76 | 0.42 | 38.74 | 4.76 |
| 5  | 0.83 | 36.32 | 6.02 | 0.83 | 36.32 | 6.02 |
| 6  | 0.00 | 31.32 | 3.35 | 0.00 | 31.44 | 3.52 |
| 7  | 0.00 | 28.44 | 2.90 | 0.00 | 28.64 | 3.20 |
| 8  | 0.00 | 30.35 | 3.25 | 0.00 | 30.88 | 3.94 |
| 9  | 0.33 | 28.99 | 3.85 | 0.33 | 29.10 | 4.00 |
| 10 | 0.31 | 27.26 | 2.84 | 0.31 | 27.64 | 3.31 |
| 11 | 0.00 | 24.40 | 2.26 | 0.31 | 24.57 | 2.47 |
| 12 | 0.00 | 25.36 | 2.80 | 0.00 | 26.19 | 3.87 |
| 13 | 0.00 | 21.52 | 1.54 | 0.31 | 22.50 | 2.72 |
| 14 | 0.00 | 20.99 | 2.83 | 0.00 | 21.32 | 3.24 |
| 15 | 0.56 | 22.77 | 4.23 | 0.56 | 22.91 | 4.41 |
| 16 | 0.00 | 22.48 | 3.00 | 0.00 | 22.90 | 3.51 |

We see that the improvements of the DC values are very small. Indeed, the heuristic of Engel with the "Kali" rule is known to give near-optimal results for lexmin(DT,DC) on random instances [14]. On the other hand, the improvements of the $SU_{var}$ values are remarkable. In comparison with the values obtained by the heuristic of Engel with the "Kali" rule and LKH, we obtain improvements from 20% to 43%. Comparing to the heuristic of Engel with the "Last" rule and LKH (which is one of the initial solution of 2PPLS), we obtain improvements from 1% to 6%.

We also see that allowing to deteriorate the DT value is only interesting for some instances. Both objectives DC and $SU_{var}$ seem thus positively correlated with the DT objective.

### 5.3.2 Real instances

Similar results have been found for the real instances. Here are the main conclusions:

- The number of potentially efficient solutions is still low: between 1 and 6.
- The number of phases is between 2 and 13.
- On the other hand, the running time is higher, between 3 and 1125 seconds. Indeed, the size of some of the instances is higher than the size of the random instances.

- The DC values comparing to the heuristic of Engel have only been improved for 5 instances out of 25.
- On the other hand, for $SU_{var}$, comparing to the values obtained by the heuristic of Engel with the "Kali" rule and LKH, we obtain improvements from 13% to 37%. Comparing to the heuristic of Engel with the "Last" rule and LKH, we obtain improvements from 0.65% to 15%. The DC and $SU_{var}$ objectives are still positively correlated with the DT objective, since the results with DT not necessary optimal are very similar to those obtained if DT is optimal.

We give in Table 5 the values of the objectives for the six non-dominated points found for a real instance of size 23x20. We see that is possible to reduce quite considerably the $SU_{var}$ value if we do not use the solutions that give the optimal values for DT or DC.

Table 5
Values of the objectives for the six potentially non-dominated points found for a real instance of size 23x20.

| DT | DC | $SU_{var}$ |
|----|----|------------|
| 91 | 20 | 260 |
| 91 | 21 | 220 |
| 92 | 20 | 241 |
| 92 | 21 | 212 |
| 93 | 22 | 211 |
| 97 | 22 | 209 |

# 6 Conclusion and perspectives

We have presented in this work first results for the problem of multiobjective decomposition of nonnegative integer matrices, within the framework of the radiotherapy treatment. The results obtained with 2PPLS are encouraging since the method allows to improve state-of-the-art results.

More experiments on different types of instances (size and characteristics) will have to be carried out to obtain more information about the correlation between the objectives and to validate the approach. Actually, we have already realized some experiments to measure the correlation between the objectives, and we have obtained positive coefficients equal to 0.6 for (DT,DC) and (DC,$SU_{var}$) and a coefficient equal to 0.3 for (DT,$SU_{var}$). These results have been obtained by generating random solutions of random instances, but as the way the random instances and solutions are generated can have an impact on the values of the coefficients, the analysis should be made more thorough. These first results confirm however that the objectives are positively correlated.

For improving the solving method, we think that the key point will be to reduce the size of the neighborhood in order to be able to apply it not only from non-dominated solutions, as done in PLS. Indeed, in PLS, we do not apply too much the neighborhood since we only start it from new non-dominated solutions and this number remains low for all the instances treated. If it was not the case, the running time would be very high.

A method based on weight sets would be interesting since we have measured on the results that we have obtained that on average only 8% of the potentially non-dominated points generated were non-supported. However, the size of the neighborhood should be reduced. Indeed, the weight sets would define different directions of research, and for each direction a local search could be applied. That means that for each weight set, the neighborhood would be necessary. The running time of such a method will be thus high if the neighborhood is very time-consuming.

To reduce the size of the neighborhood, it would be necessary to deeply study what are the successful moves, what gives improvements of the solution and what never gives improvements. With such results, we will be able to reduce the size of the neighborhood and making it faster and consequently more efficient. Random choices in the neighborhood with a limited number of neighbors would also be a valuable option.

Defining a crossover operator would be interesting, and would allow to integrate the local search in an evolutionary algorithm.

We leave all these ideas for further research.

Finally, we have only considered the third problem that occurs in the IMRT optimization process, without taking into account the geometry and intensity problems (we suppose that the intensity matrices are given). Solving the three problems simultaneously would be very challenging (but as said by Ehrgott *et al.* in their recent survey [7]: "This is, however, beyond the capabilities of current optimization algorithms and computer power"!).

**Acknowledgments**

We thank Céline Engelbeen from the Université libre de Bruxelles for introducing this subject and for several discussions. T. Lust thanks the "Fonds National de la Recherche Scientifique" for a research fellow grant (Aspirant F.R.S. - FNRS).